\documentclass[twocolumn]{aastex63}

\usepackage{textcomp}
\usepackage{CJKutf8}
\usepackage{amsmath}

\shorttitle{A Gravitational Redshift Measurement of the WD M--R Relation}
\shortauthors{Chandra, Hwang, Zakamska \& Cheng}

\graphicspath{{./}{figures/}}

\begin{document}

\title{A Gravitational Redshift Measurement of the White Dwarf Mass--Radius Relation}

\author[0000-0002-0572-8012]{Vedant Chandra}
\affiliation{Department of Physics \& Astronomy, Johns Hopkins University,
3400 N Charles St,
Baltimore, MD 21218, USA}

\author[0000-0003-4250-4437]{Hsiang-Chih Hwang}
\affiliation{Department of Physics \& Astronomy, Johns Hopkins University,
3400 N Charles St,
Baltimore, MD 21218, USA}

\author[0000-0001-6100-6869]{Nadia L. Zakamska}
\affiliation{Department of Physics \& Astronomy, Johns Hopkins University,
3400 N Charles St,
Baltimore, MD 21218, USA}

\author[0000-0002-9156-7461]{Sihao Cheng \begin{CJK*}{UTF8}{gbsn}(程思浩)\end{CJK*}}
\affiliation{Department of Physics \& Astronomy, Johns Hopkins University,
3400 N Charles St,
Baltimore, MD 21218, USA}

\begin{abstract}

 The mass--radius relation of white dwarfs is largely determined by the equation of state of degenerate electrons, which causes the stellar radius to decrease as mass increases. Here we observationally measure this relation using the gravitational redshift effect, a prediction of general relativity that depends on the ratio between stellar mass and radius. Using observations of over three thousand white dwarfs from the Sloan Digital Sky Survey and the \textit{Gaia} space observatory, we derive apparent radial velocities from absorption lines, stellar radii from photometry and parallaxes, and surface gravities by fitting atmospheric models to spectra. By averaging the apparent radial velocities of white dwarfs with similar radii and, independently, surface gravities, we cancel out random Doppler shifts and measure the underlying gravitational redshift. Using these results, we empirically measure the white dwarf mass--radius relation across a wide range of stellar masses. Our results are consistent with leading theoretical models, and our methods could be used with future observations to empirically constrain white dwarf core composition and evolution.

\end{abstract}


\section{Introduction}

White dwarfs, the end--life stages of nearly all stars in the Universe, are fascinating examples of quantum-mechanical effects on a macroscopic scale. They have masses similar to the Sun but radii similar to the Earth, and are consequently extremely dense. White dwarfs are formed from the hot stripped cores of stars that have exhausted their nuclear fuel. They spend the rest of their lives slowly cooling via radiation, with typical surface temperatures ranging between $4\ 000-100\ 000$ K, and typical masses ranging between $0.2-1.35\ M_\odot$. 

To leading order, white dwarfs are supported against their own gravity by the pressure of degenerate electron gas, resulting in a mass--radius relationship such that radius decreases as mass increases \citep{Chandrasekhar1933}. This mass--radius relation is governed by the equation of state (EoS) of white dwarf material, and it has a multitude of astrophysical ramifications. In Galactic archaeology, cooling white dwarfs can provide an independent age measurement of a stellar population and set a lower limit on the age of the Universe \citep{Hansen2002}. In cosmology, the EoS is crucial to our understanding of type Ia supernova progenitors and their explosion mechanisms \citep{Maoz2014}. Directly measuring the white dwarf mass--radius relation and, through it, the equation of state, is an outstanding astrophysical problem \citep{Tremblay2017}.

The physical parameters of white dwarfs can be measured from photometry or spectroscopy, or from a combination of the two. The effective surface temperature $T_{\rm eff}$ can be measured from broad-band photometry. With this in hand, if the trigonometric distance to a white dwarf is known, its apparent solid angle and consequently its radius can be calculated (hereafter the `photometric radius'). Furthermore, on white dwarf spectra, the shape of photospheric absorption lines depends on physical conditions like pressure and effective temperature. Therefore, by fitting model atmospheres to white dwarf spectra, one can measure the best-fitting effective surface temperature $T_{\rm eff}$ and surface gravity $\log{g}$ (hereafter `spectroscopic $\log{g}$', with $g$ in cgs units of $\rm cm\ s^{-2}$). 

Several studies have compared photometric and spectroscopic observables for white dwarfs \citep{Bedard2017,Joyce2018b,Tremblay2019,GB2019}. Most of these studies assumed a theoretical mass--radius relation \citep{Fontaine2001} that has repeatedly been shown to be broadly consistent with observations. A more complete summary of other `semi-empirical' tests of the mass--radius relation is provided in \cite{Tremblay2017}. 

In order to directly test the mass--radius relation itself, an additional constraint on mass and radius is provided by the gravitational redshift. This effect follows from the equivalence principle of general relativity and causes photons originating from gravitational potentials (like the surface of a white dwarf) to be shifted to redder wavelengths \citep{Einstein1916}. This causes a measurable wavelength shift in photospheric absorption lines that is proportional to the ratio between the stellar mass and radius, and is entirely independent of theoretical models. However, the gravitational redshift is difficult to isolate since the absorption lines are also Doppler shifted due to random stellar motions along the line of sight. 

The degeneracy between Doppler shift and gravitational redshift can be overcome by using other information like co-moving companions to constrain the radial velocity component of the apparent velocity, as was done in the case of Sirius B \citep{Joyce2018}, as well as for white dwarfs in the Hyades cluster \citep{Pasquini2019}. These kinds of studies require high resolution spectroscopy to accurately measure individual apparent radial velocities, and have consequently probed narrow mass ranges and had small sample sizes. Regardless, they have shown broad agreement with predictions from the theoretical mass--radius relation \citep{Joyce2018b, Romero2019}. One of the most quantitative and model-independent tests of the mass-radius relation was presented by \cite{Parsons2017}, who studied 16 white dwarfs in detached eclipsing binaries and found similar agreement between observations and theory.

Alternatively, for a sample of field white dwarfs, it is possible to average the observed apparent radial velocities to cancel out the Doppler effect from random motions relative to the Sun, and hence measure the mean gravitational redshift. This method was pioneered by \cite{Falcon2010}, who averaged the apparent radial velocities of 449 stars to derive the mean mass of hydrogen-atmosphere white dwarfs using an assumed mass--radius relation from \cite{Fontaine2001}. 

In this work, we apply the method of averaged gravitational redshifts to a much larger sample of stars without assuming any mass--radius relation. We use \textit{Gaia} astrometry and SDSS spectro-photometry to derive apparent radial velocities, photometric radii, and spectroscopic $\log{g}$ for a sample of over three thousand DA (hydrogen-rich atmosphere) white dwarf stars. We statistically uncover the dependence of gravitational redshift on stellar radius and surface gravity, and compare our results to predictions from the theoretical mass--radius relation of \cite{Fontaine2001}. To our knowledge this is the first empirical measurement of the white dwarf mass--radius relation across a wide range of masses and with such a large sample of stars.

We describe our radial velocity, photometric radius, and spectroscopic $\log{g}$ measurements in Section \ref{sec:analysis} along with details about our sample selection. We discuss possible biases in our redshift and radius measurements in Section \ref{sec:systematic} and develop methods to debias our results. We present our main results of gravitational redshift as a function of radius and as a function of $\log{g}$ in Section \ref{sec:results}, and we discuss our findings and their implications in Section \ref{sec:discussion}. 

\section{Data Analysis}\label{sec:analysis}

\subsection{Sample Selection}

Our parent sample consists of 20,088 spectroscopically-confirmed white dwarfs \citep{Kepler2019} from the Sloan Digital Sky Survey (SDSS; \citealt{SDSS2017}). We build our sample of DA white dwarfs by applying selection cuts based on SDSS and \textit{Gaia} data. We begin by selecting stars with a `DA' classification from \cite{Kepler2019}. To enable the accurate measurement of absorption line centroids (and hence apparent radial velocities), we apply a selection cut of signal-to-noise ratio (S/N) $\geq$ 10 and visually confirm that all stars in the sample are indeed DA white dwarfs (strong Balmer lines on an otherwise featureless spectrum). This leaves us with 7184 DA white dwarf spectra for which we derive our apparent radial velocity $v_{\rm app}$, photometric radius, and spectroscopic $\log{g}$ measurements. We then apply further cleanliness cuts on the basis of our measurement uncertainties. We use separate cuts for our redshift--radius and redshift--$\log{g}$ results since they rely on different measurements.
 
For the redshift--radius results, our photometric radius measurements are sensitive to the \textit{Gaia} parallaxes $\pi$, since they assume a known distance to the star. We therefore select stars with $\pi / \sigma_\pi > 10$. We also apply a cut on radial velocity uncertainties such that $\sigma(v_{\rm app}) < 50\ \rm km\ s^{-1}$. We find that adopting a more stringent cut, say $\sigma(v_{\rm app}) < 25\ \rm km\ s^{-1}$, halves our sample size and makes our final statistical result in Figure \ref{fig:massvrv_fontaine} noisier but qualitatively unchanged. We therefore elect to adopt a more relaxed selection cut on individual radial velocities to maintain a larger sample size and more effectively cancel out random stellar motion during the averaging process. Finally, we restrict our sample to $d \leq 500$ parsecs -- where the distance $d = 1/\pi$ -- to assume a locally co-moving population. After applying these selection cuts, we have a sample size of $3316$ stars for our photometric results in Section \ref{photresult}. 

For the redshift--$\log{g}$ results we apply the same cuts on radial velocity uncertainty and distance, $\sigma(v_{\rm app}) < 50\ \rm km\ s^{-1}$ and $d < 500\ \rm pc$. To remove bad spectroscopic fits with spurious $\log{g}$ measurements (Section \ref{sec:speclogg}), we select stars with a reduced chi-square discrepancy between the observed and synthetic spectra $\chi_r^2 \leq 1$. We find that most of the stars we remove with $\chi_r^2 > 1$ either have noisy spectra or deviations from the DA spectrum (eg. weak magnetism) that made it past our initial selection cuts. There are not enough stars with $\log{g} < 7.6$ or $ \log{g} > 9$ to satisfy our requirement of having $> 50$ stars per $\log{g}$ bin. We therefore restrict our analysis to stars with $7.6 \leq \log{g} \leq 9$. We end up with a sample of $2577$ stars for our spectroscopic results in Section \ref{specresult}.

\subsection{Redshift Measurements} \label{redshifts}

The theory of general relativity predicts that photons redshift as they climb out of a gravitational potential \citep{Einstein1916}. Therefore, photons originating from quantum-mechanical transitions in gravitational potentials are observed at longer wavelengths than theoretically expected. The atmospheres of DA white dwarfs produce strong absorption lines in the hydrogen Balmer series. Due to the gravitational redshift from the surfaces of white dwarfs, the central wavelength of these lines is shifted by an amount $\delta \lambda=\lambda_0 v_g/c$, where $\lambda_0$ is the laboratory wavelength of the line and
\begin{equation}\label{gredshift}
v_g = \frac{\delta \lambda \cdot c}{\lambda_0} =\frac{GM}{Rc} \end{equation}
is the recession velocity that would result in an equivalent redshift. Thus, gravitational redshift is proportional to $M/R$ and is a sensitive and model-independent probe of the white dwarf mass--radius relation.

We start by measuring the apparent radial velocity of each star $v_{\rm app}$, which is the sum of the gravitational redshift and the relative motion between white dwarf and the Sun. There are several challenges when measuring high-precision redshifts from white dwarf spectra, mostly due to their asymmetrically pressure-broadened absorption lines. In studies with high-resolution spectroscopy \citep{Falcon2010}, the narrow non-broadened cores of hydrogen Balmer lines ($\sigma_v \simeq 30\ \rm km\ s^{-1}$) are used to accurately derive the redshift. However, SDSS spectra are resolution-limited ($\sigma_v \simeq 60\ \rm km\ s^{-1}$ at H$\rm \alpha$) and the line core itself is somewhat blended with the pressure-broadened wings. \cite{Joyce2018} measured the gravitational redshift of Sirius B using the mid-resolution ($\sigma_v \simeq 25\ \rm km\ s^{-1}$ at H$\rm \alpha$) STIS spectrograph on the Hubble Space Telescope, which also barely resolved the line core. We adapt the iterative fitting technique from \cite{Joyce2018} to produce unbiased redshift measurements from our mid-resolution SDSS spectra. 

\begin{figure}
    \centering
    \includegraphics[width=\columnwidth]{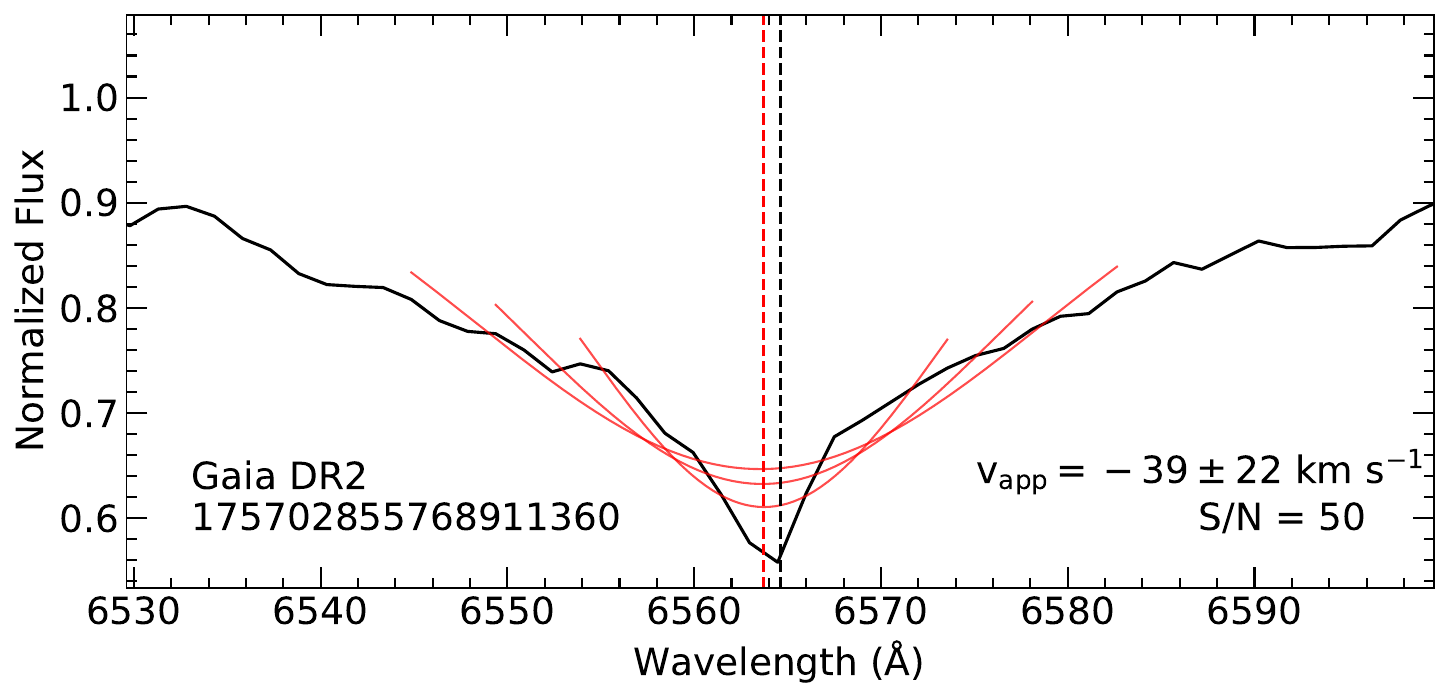}
    \includegraphics[width=\columnwidth]{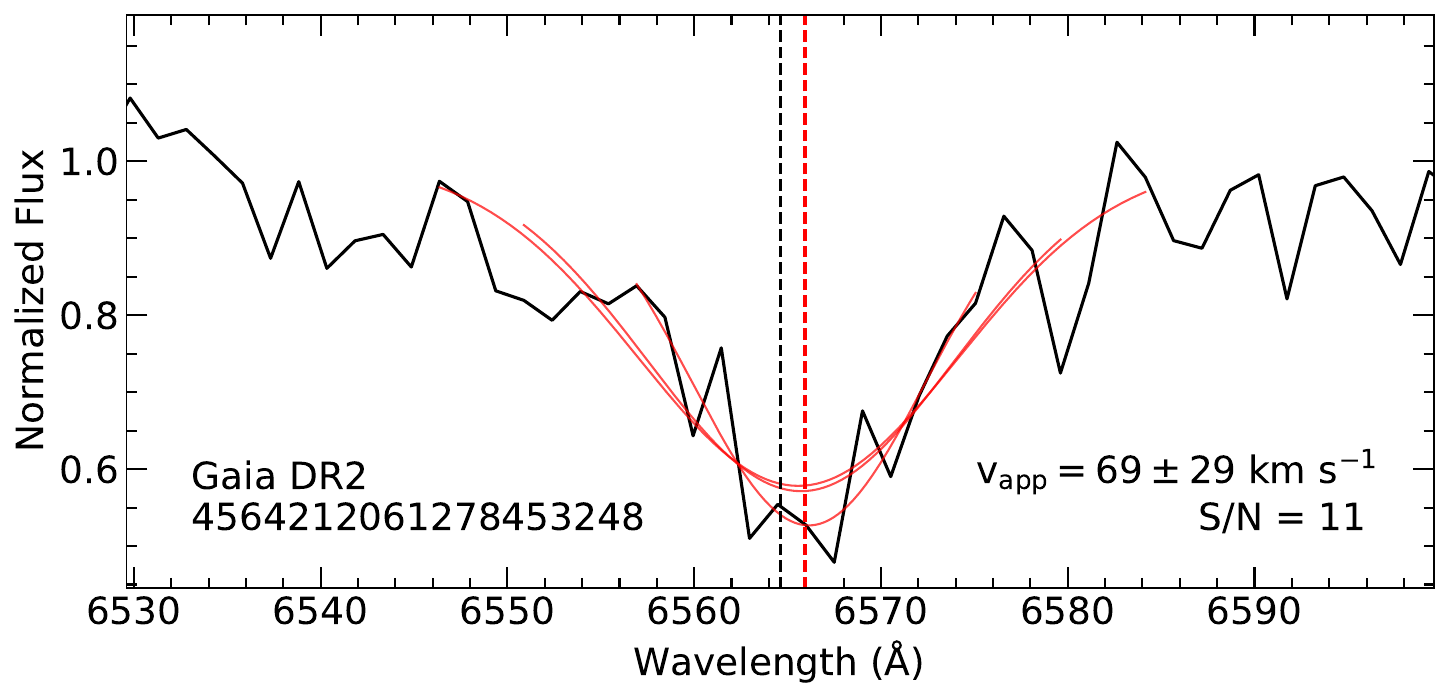}
    \caption{Our procedure to fit absorption lines for apparent radial velocity. We fit the H$\alpha$ line for two stars in our sample with high (top) and low (bottom) signal-to-noise ratios respectively. The rest-frame transition wavelength is indicated by a black dashed line, and our final fitted centroid by a red dashed line.}
    \label{fig:linefits}
\end{figure}

We fit line profiles to successively cropped windows (from $25 - 10$ \AA\ wide in $5$ \AA\ steps) around the H$\alpha$ absorption line core (Figure \ref{fig:linefits}). In this wavelength regime within 25 \rm{\AA}\ of the line core, the asymmetric Stark shift is below 10 $\rm km\ s^{-1}$ for even the most massive stars \citep{Halenka2015, Joyce2018}. Avoiding Stark bias in the H$\beta$ and H$\gamma$ lines would require us to fit such a small window of wavelengths that we would be restricted to only the highest-S/N spectra, which would drastically reduce our sample size. We therefore use only the H$\alpha$ absorption line for our analysis.

We use the {\sc lmfit} package in Python to fit the line profiles with a non-linear least squares algorithm \citep{LM1978, Newville2014}. We report the centroid of the final fitted profile with the 10\ \AA\ window as our measurement. We derive the uncertainty of this quantity as the dispersion (standard deviation) between the measurements across all the windows, capturing the statistical uncertainty introduced by considering a greater or lesser number of pixels in the fit. To this, we add in quadrature the uncertainty reported by the covariance matrix of our fitting routine. We convert observed line centroids to apparent radial velocities with
\begin{equation}
    v_{\rm app} = \left(\frac{\lambda_{\mathrm{obs}} - \lambda_0}{\lambda_0}\right)\cdot c
\end{equation} where $\lambda_{obs}$ is the observed wavelength, $\lambda_0$ is the rest-frame wavelength of the absorption line and $c$ is the speed of light. 

\begin{figure}
    \centering
    \includegraphics[width=\columnwidth]{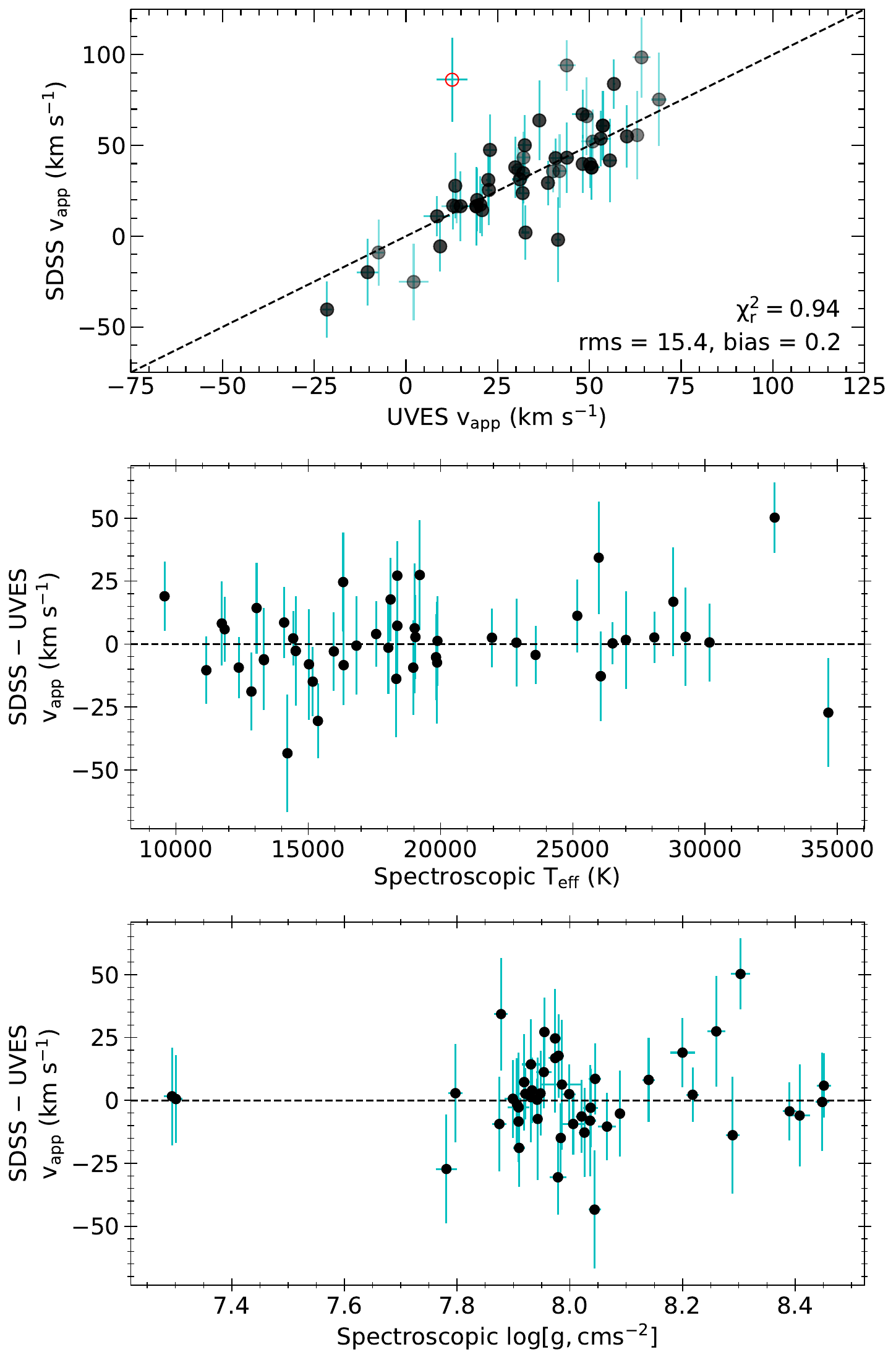}
    \caption{Comparison of LSR-corrected apparent radial velocities measured in this study using mid-resolution SDSS spectra to those obtained using high-resolution UVES spectra \citep{Falcon2010}. Top: the dashed line represents the 1-1 correlation. The outlier with an unfilled circle has a mass of 0.32 $M_\odot$ and is therefore likely in a binary system with a less luminous, more massive companion, explaining the discrepant radial velocity compared to the older measurement. Middle and bottom: residual difference as a function of spectroscopic temperature and surface gravity respectively.}
    \label{fig:falcon_compare}
\end{figure}

We validate our radial velocity measurements by comparing them to prior results obtained on the same targets with higher-resolution spectroscopy. \cite{Falcon2010} measured the apparent radial velocity of 449 hydrogen atmosphere white dwarfs by directly fitting the line cores of high-resolution ($\sigma_v = 16\ \rm km\ s^{-1}$ at H$\rm\alpha$) optical spectra. We identify 86 shared targets between our samples (Figure \ref{fig:falcon_compare}, top) and find a root mean squared velocity difference of 15.4 $\rm km\ s^{-1}$ with a mean difference $< 1\ \rm km\ s^{-1}$. We also verify that our velocity measurements are unbiased as a function of spectroscopic $T_{\rm eff}$ and $\log{g}$ (Figure \ref{fig:falcon_compare}, middle and bottom). Section \ref{sec:speclogg} has more details about our spectroscopic measurements of $T_{\rm eff}$ and $\log{g}$. 

Another way to validate our radial velocity measurements is to perform mock measurements on theoretical spectra. We perform Monte Carlo tests on synthetic spectra of hydrogen atmosphere white dwarfs \citep{Koester2010}, smoothing and down-sampling them to the resolution of SDSS spectra. We test our fitting routine on synthetic spectra across a range of temperatures and surface gravities, and find that our apparent radial velocities are negligibly biased by Stark asymmetry except for cool white dwarfs with a very high surface gravity. In our sample, the surface gravity distribution is sharply peaked around $\log{g} = 8$, and $\log{g}$ is $< 9$ for all stars.

We find that with our chosen window sizes and fitting routine for the H$\alpha$ line, the bias in radial velocity due to Stark asymmetry \citep{Tremblay2009,Halenka2015} is $< 10\ \rm km\ s^{-1}$ for all temperatures and surface gravities, and far lower for the majority of temperatures and surface gravities present in our sample. We conclude that our apparent radial velocity measurements from medium-resolution SDSS spectra are noisy but unbiased.

 SDSS spectra are wavelength-calibrated in the solar system barycenter. We assume our sample is a locally co-moving population since all stars lie within 500 parsecs of the Sun \citep{Binney1987}. Under this assumption, we correct our apparent radial velocity measurements to the local standard of rest (LSR). We assume the solar velocity relative to the local standard of rest to be $(U,V,W)_\odot = (11.1, 12.24, 7.25)\ \rm km\ s^{-1}$ \citep{Schonrich2010}. 
 
 We use the coordinate transform tool from {\sc astropy} to perform a 3-D velocity transform using our measured radial velocities along with geometric distances from \textit{Gaia} \citep{BJ2018} and calculate LSR-corrected radial velocities for our sample. We discuss the effects of this transformation further in Section \ref{kinematics}. After this correction, the observed apparent radial velocity is solely due to the gravitational redshift plus the Doppler shift of random stellar motions caused by non-circular orbits in the galaxy \citep{Falcon2010}.
 
 \vspace{12pt}

\subsection{Photometric Radii}

The gravitational redshift is a direct measurement of the ratio between white dwarf mass and radius $M/R$ (Equation \ref{gredshift}). To probe the mass--radius relation, another independent constraint of $M$ or $R$ is needed. The so-called photometric method is well-established in the white dwarf literature (recently, \citealt{GB2019}) as a way to measure stellar radius given a spectral energy distribution (SED) and a distance to the star. We use the photometric method to derive stellar radii for all white dwarfs in our sample with multiband \textit{ugriz} SDSS photometry \citep{SDSS2017} and trigonometric parallaxes from the \textit{Gaia}  space observatory \citep{Gaia2018}. 

For a particular wavelength and stellar temperature and surface gravity, the apparent flux $f_\nu$ and flux on the white dwarf surface $H_\nu(T_\text{eff}, \log{g})$ are related by \begin{equation}\label{photmethod}
    f_\nu = 4 \pi [R/D]^2 \cdot H_\nu(T_\text{eff}, \log{g})
\end{equation} where $D$ is the distance to the star and $R$ is its radius. We derive the surface fluxes for the SDSS passbands using the latest version of the synthetic color table\footnote{\url{http://www.astro.umontreal.ca/~bergeron/CoolingModels/}} described in \cite{Holberg2006}. More details about the underlying synthetic atmospheres can be found in \cite{Tremblay2011} and references within. Rather than simply inverting the \textit{Gaia} parallaxes, we use the Bayesian geometric distances from \cite{BJ2018} as estimates of the distance $D$ for all stars in our sample. These distances better incorporate the non-linear relation between parallax and distance in their uncertainty estimation, which is helpful during our selection cuts to remove spurious radius measurements. They also incorporate a distance prior based on a Galactic length scale model to compute the posterior distance distribution for each star.

Interstellar dust in the solar neighborhood extincts and reddens incoming light from our target stars, affecting the SDSS spectral energy distribution. Since a significant portion of our sample is further than 100 pc away from the Sun, the effect of extinction cannot be ignored. We compute the \textit{ugriz} extinction for each star in our sample using the {\sc mwdust} utility \citep{Bovy2016}. For each star, we integrate combined dust maps from \cite{Marshall2006} and \cite{Green2019} using Galactic coordinates $l,b$ from \textit{Gaia} astrometry and the geometric distance $D$ from \cite{BJ2018}. We correct the observed $ugriz$ magnitudes of our white dwarfs for extinction and then perform a transformation to the AB magnitude system \citep{Eisenstein2006} with the mapping \begin{equation}
\begin{split}
    u = u_{\rm SDSS} - 0.040 \\
    g = g_{\rm SDSS} \\
    r = r_{\rm SDSS} \\
    i = i_{\rm SDSS} + 0.015 \\
    z = z_{\rm SDSS} + 0.030
    \end{split}
\end{equation}

\begin{figure}
    \centering
    \includegraphics[width=\columnwidth]{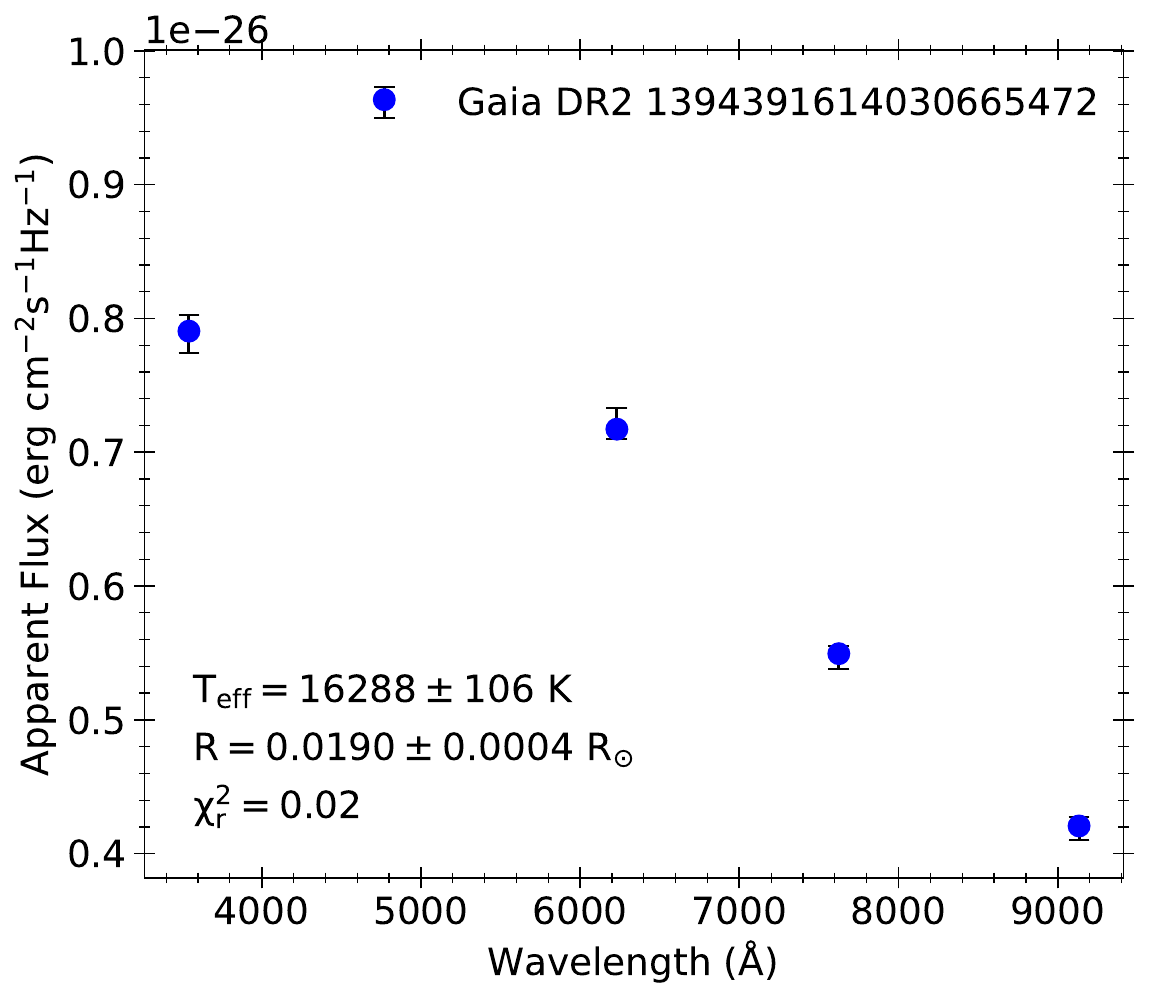}
    \caption{Example of the photometric method to derive stellar radius. The vertical error bars are derived from SDSS magnitude errors, and the uncertainties on the fitted parameters are the statistical uncertainties returned by our fitting routine's covariance matrix. The blue points indicate the best-fitting model SED.}
    \label{fig:sed_radius}
\end{figure}

We convert the de-reddened \textit{ugriz} magnitudes of observed white dwarfs and synthetic \textit{ugriz} colors generated from the synthetic models into average fluxes using the appropriate zero point flux value for the AB system. We remove the dependence of the synthetic apparent flux on the theoretical mass--radius relation by normalizing them to the surface flux $H_\nu$. We allow $T_{\rm eff}$, $\log{g}$, and $R$ to vary as free parameters, minimizing the $\chi^2$ difference between the apparent model flux and observed flux across the \textit{ugriz} SED (Figure \ref{fig:sed_radius}). 

The observed flux from a star is determined by the surface temperature and stellar radius (Equation \ref{photmethod}). The shape of the SED strongly constrains the temperature, and hence the strength of the flux provides a good measurement of $R$. However, since the strong absorption lines on the white dwarf spectrum vary as a function of $T_{\rm eff}$ and $\log{g}$, the shape of the SED has a weak dependence on $\log{g}$ as well. Therefore, although we do not use the photometric $\log{g}$ value in our analysis, we allow it to vary as a free nuisance parameter during the SED fitting procedure. We find that the alternative approach of fixing $\log{g} = 8$ does not significantly affect our main results, but leads to slightly worse-fitting SEDs.  

We use the covariance matrix returned by the Levenberg--Marquardt algorithm in {\sc lmfit} as an estimate of the radius uncertainty from photometric errors. We also compute radii using the upper and lower 1-$\sigma$ confidence bound of the geometric distance from \cite{BJ2018} and take the difference between the two as an estimate of the radius uncertainty introduced by parallax errors. We add these two uncertainties in quadrature to derive our total radius uncertainties, which are usually dominated by the parallax errors. By adding these uncertainties we are implicitly assuming that the distance confidence interval is symmetric, which it usually is not. However, we find this approximation to be sufficiently accurate for use in our selection cuts. 

\subsection{Spectroscopic Surface Gravity} \label{sec:speclogg}

Apart from the photometric radius, another independent observable is the spectroscopic surface gravity $\log{g}$. Surface gravity $g = GM/R^2$ is a function of both mass and radius, and therefore the white dwarf mass-radius relation is encoded in $\log{g}$. Surface gravity, in turn, is imprinted on the stellar spectrum via the pressure broadening of absorption lines \citep{Tremblay2009}. Spectroscopic $\log{g}$ is not as direct an observable as photometric radius and gravitational redshift, because it is more sensitive to the atmospheric models \citep{Koester2010}. Nevertheless, recent studies \citep{Joyce2018b,Tremblay2019,Bergeron2019,GB2019} have demonstrated its accuracy and consistency. 

The so-called spectroscopic method is a standard technique to derive $T_{\rm eff}$ and $\log{g}$ by comparing observed spectra to synthetic spectra generated by atmospheric models. Since all stars in our sample are spectroscopically-confirmed DA white dwarfs, we use a grid of pure-hydrogen atmosphere synthetic spectra from \cite{Koester2010}. These 1-D models assume local thermodynamic equilibrium (LTE) and incorporate the non-ideal hydrogen Stark profiles from \cite{Tremblay2009}. The grid of synthetic spectra\footnote{\url{http://svo2.cab.inta-csic.es/theory/newov2/index.php}} spans $6000\ \mathrm{K} \leq T_{\rm eff} \leq 40000$ K and $6.5 \leq \log{g} \leq 9.5$, with $g$ in cgs units of $\rm cm\ s^{-2}$.
 
 We use the {\sc wdtools} utility \citep{Chandra2020} in Python to interpolate these model spectra in wavelength, $T_{\rm eff}$, and $\log{g}$. We convolve the synthetic spectra to the SDSS resolution with a 3 \AA\ Gaussian kernel, and continuum-normalize hydrogen Balmer lines of the observed and synthetic spectra from H$\alpha$ to H$8$. For each star, we perform two independent fits using {\sc lmfit} to minimize the $\chi^2$ difference between the observed and synthetic spectra, one with a cool prior ($6000\ \mathrm{K} \leq T_{\rm eff} \leq 15000\ \mathrm{K}$) and one with a warm prior ($15000\ \mathrm{K} < T_{\rm eff} \leq 40000\ \mathrm{K}$). We select parameters from the fit with a lower $\chi^2$. This alleviates a known problem with fitting white dwarf spectra where degenerate $\chi^2$ minima exist for both hot and cold solutions. We could resolve this degeneracy using our photometry and parallaxes, but we elect to keep our spectroscopic fitting procedure independent of photometric information. 
 
 \begin{figure}
     \centering
     \includegraphics[width=\columnwidth]{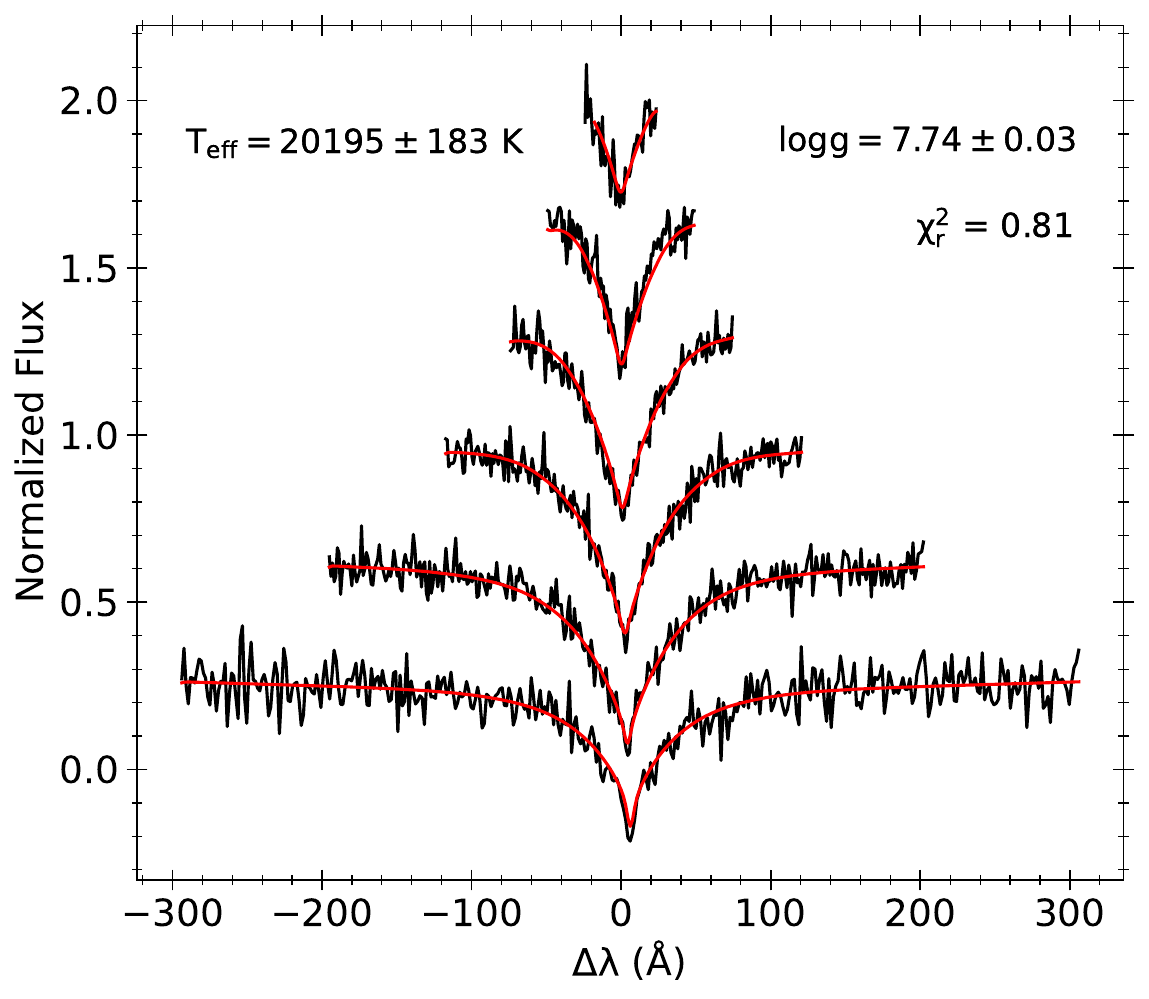}
     \caption{Example of our spectroscopic fitting procedure to derive $\log{g}$. In black are the continuum-normalized Balmer lines of a random star in our sample (SDSS J223656.92+323820.8) and in red is the best-fitting synthetic spectrum. The Balmer lines are arranged H$\alpha$--H$8$ from bottom to top.}
     \label{fig:gfp_fit}
 \end{figure}
 
 Throughout the fitting procedure, we allow the wavelength shift between the synthetic and observed spectra to vary as a free nuisance parameter. In principle, this parameter provides another estimate of the apparent radial velocity of the star. However, this measurement can be biased by systematic differences in the template synthetic spectra and the observed spectra. We therefore use the simpler and model-independent methodology of Section \ref{redshifts} to derive apparent radial velocities.
 
 For each star we derive uncertainties for $T_{\rm eff}$ and $\log{g}$ using the {\sc emcee} Markov Chain Monte Carlo sampler \citep{FM2019}, initializing 250 walkers around our initial best-fit parameters and sampling for 100 steps to burn-in, then sampling another 100 steps to get samples from the posterior distributions of $T_{\rm eff}$ and $\log{g}$.  We select the sample with the lowest $\chi^2$ as our solution for $T_{\rm eff}$ and $\log{g}$, and compute the standard deviation of each marginalized posterior distribution to get 1-$\sigma$ uncertainties (a sample fit is shown in Figure \ref{fig:gfp_fit}). These uncertainties are purely statistical in nature, and we use them to filter out ill-constrained $\log{g}$ values in our quality selection cuts. There are uncertainties in the physics behind the spectroscopic templates we use, and hence the systematic uncertainties in our measurements can be up to 2\% in $T_{\rm eff}$ and $0.2\ \rm dex$ in $\log{g}$ \citep{Tremblay2019}. 
 
Whilst past studies \citep{Kepler2019, Tremblay2019} have already derived spectroscopic $\log{g}$ for most stars in our sample using their own model atmospheres, we elect to re-compute $\log{g}$ using the methods described above. Most previous spectroscopic studies utilize prior knowledge of effective temperature obtained via fits to photometric data, which implicitly make use of the mass-radius relation. Since we want our spectroscopic results to be as independent from our photometric results as possible, we compute our own values for $\log{g}$. As a check, we confirm that our fitted $\log{g}$ values are broadly consistent with those from \cite{Kepler2019}, finding a root-mean-square difference of 0.15 dex. This is comparable to the systematic uncertainty introduced by fitting $\log{g}$ using atmospheric models from different groups \citep{Tremblay2019}, as is the case here.

\section{Systematic Effects} \label{sec:systematic}

\subsection{Non-Uniform Distribution of Radii} \label{debias}
 
The white dwarf mass distribution \citep{Kilic2018} is sharply peaked around $0.6\ M_{\odot}$ -- consequently, the radius distribution is sharply peaked at $\sim 0.012\ R_\odot$. There is hence a high probability for an average radius white dwarf to contaminate the low- and high-radius bins of our results in Figure \ref{fig:massvrv_fontaine} (left) due to random errors, flattening the expected relation. This is analogous to the bias discussed in \cite{Lauer2007}.

We investigate the impact of this effect with Monte Carlo simulations.  We sample `noiseless' radii by converting a theoretical white dwarf mass distribution \citep{Kilic2018} -- a Cauchy distribution centred at 0.6 $M_{\odot}$ with $\gamma = 0.05\ M_{\odot}$ -- into a distribution of radii and generate corresponding theoretical gravitational redshifts using theoretical models \citep{Fontaine2001} and Equation \ref{gredshift}. We add Gaussian noise $\rm \delta R$ to the radius samples and repeat the analysis of Figure \ref{fig:massvrv_fontaine} with these synthetic datasets. Radius errors on the order of the median radius uncertainty of our sample ($\delta R \sim 0.002\ R_\odot$) clearly flatten the expected relation, biasing the outer bins by as much as 15 $\rm km s^{-1}$. We present our results with and without a correction of this known bias. 

\subsection{Unresolved Low-Mass Binaries}

Another possible bias in our photometric radii is the presence of unresolved stellar binaries. There is strong evidence to suggest that nearly all white dwarfs with mass $\leq 0.45\ M_{\odot}$ exist in binary systems \citep{Brown2016}, since the main-sequence evolution time of nominal progenitors of these stars is greater than the age of the Universe. This rules out single-star evolution and suggests that all these white dwarfs are formed by binary evolution via stable Roche lobe overflow or unstable mass transfer \citep{Sun2018}. White dwarfs with main sequence companions have been spectroscopically excluded from our sample in the initial selection cuts \citep{Kepler2019}. Therefore, it is safe to assume that white dwarfs with measured masses $\lesssim 0.45\ M_{\odot}$ are double-degenerate and have a more massive, more compact companion. The unresolved companion white dwarfs cause us to overestimate luminosity, which translates into an over-estimated photometric radius. This effect can bump stars from mid-radius bins into high-radius bins and flatten the expected relation. 

We quantify this bias with a Monte Carlo simulation, again using the theoretical mass distribution from \cite{Kilic2018}. We generate a synthetic sample of masses drawn from this distribution, and temperatures drawn from our empirical photometric temperature distribution and convert them into synthetic SDSS colors. For the stars with mass $\leq 0.45\ M_{\odot}$, we draw companion masses \citep{Brown2016} from a Gaussian centered at $0.76\ M_\odot$ and with a dispersion $0.25\ M_\odot$, and add the two fluxes to simulate the binary over-luminosity. For stars with mass $> 0.45\ M_{\odot}$, we assume a double-degenerate binary fraction of 1\% and simulate binary over-luminosity for that fraction as well. We convert the binary-contaminated SEDs back into photometric radius estimates, and repeat this paper's analysis on these simulated measurements. 

As expected, the binary contamination causes an over-density of high-radius measurements, which corresponds to a slight flattening in the derived redshift--radius relation. We account for this effect in our final debiased results.

\subsection{Galactic Kinematics}\label{kinematics}

Our targets in the Sloan Digital Sky Survey are not uniformly distributed across the sky. We must therefore investigate asymmetric effects like differential Galactic rotation and asymmetric drift to test our assumption of a locally co-moving population. As a start, we verify that our sample is locally co-moving with the Sun and has a zero net velocity relative to the Sun. We split our sample by Galactic coordinates into outer ($90 < l < 270$), inner ($l < 90, l > 270$), forward ($l < 180$), behind ($l > 180$), above ($b > 0$) and below ($b < 0$) subsamples. 

We compute the mean LSR-corrected radial velocity for each subsample, and compute the relevant pairwise differences to derive net $\Delta (U, V, W)$ velocities relative to the Sun (for example, $\Delta V = \langle v_r \rangle_{\rm forward} - \langle v_r \rangle_{\rm behind}$). We calculate $\Delta (U,V,W) = (0.3 \pm 1.8, -8.6 \pm 1.9, 4.4 \pm 2.3)\ \rm km\ s^{-1}$. Within our uncertainties, this is consistent with the expected asymmetric drift for this population, since we restrict our sample to within 500 parsecs of the Sun. 

We also test the robustness of our sample to a different parameterization of the local standard of rest (LSR). For all results in this paper we assume a solar velocity relative to the LSR of $(U,V,W)_\odot = (11.1, 12.24, 7.25)\ \rm km\ s^{-1}$ from \cite{Schonrich2010}. Another study \citep{Bovy2012} derived a different solar motion with a particularly higher $V_{\odot}$. We repeat the above test with $(U,V,W)_\odot = (9, 20, 10)\ \rm km\ s^{-1}$, and obtain $\Delta (U,V,W) = (-0.4 \pm 1.8, -4.6 \pm 1.8, 4.7 \pm 2.3)\ \rm km\ s^{-1}$ for our sample, once again consistent with the expected asymmetric drift. Therefore, our assumption of a locally co-moving sample holds and is robust to the choice of LSR.

\section{Results}\label{sec:results}

\begin{figure*}
    \centering
    \includegraphics[width=\columnwidth]{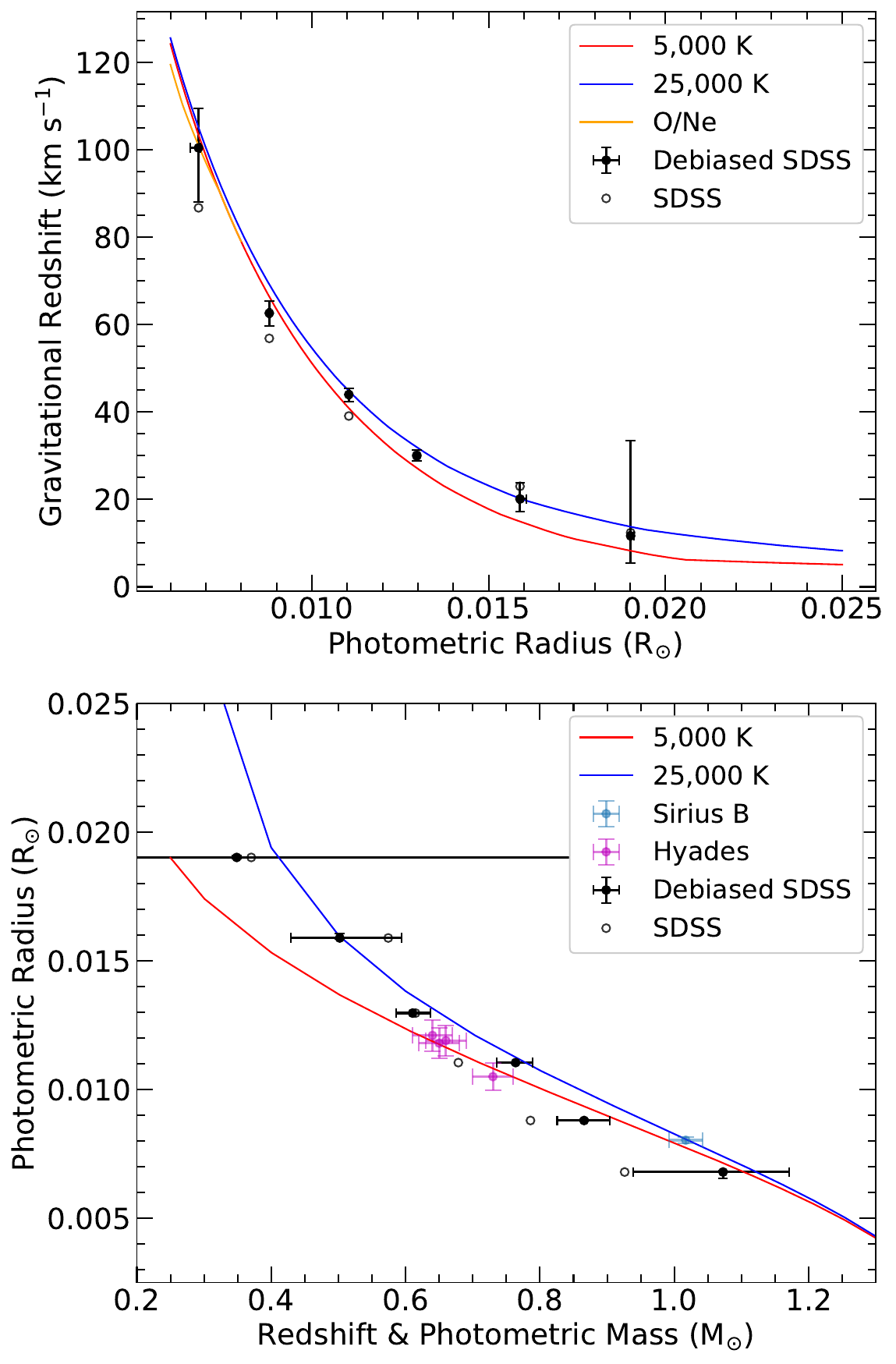}
    \includegraphics[width=\columnwidth]{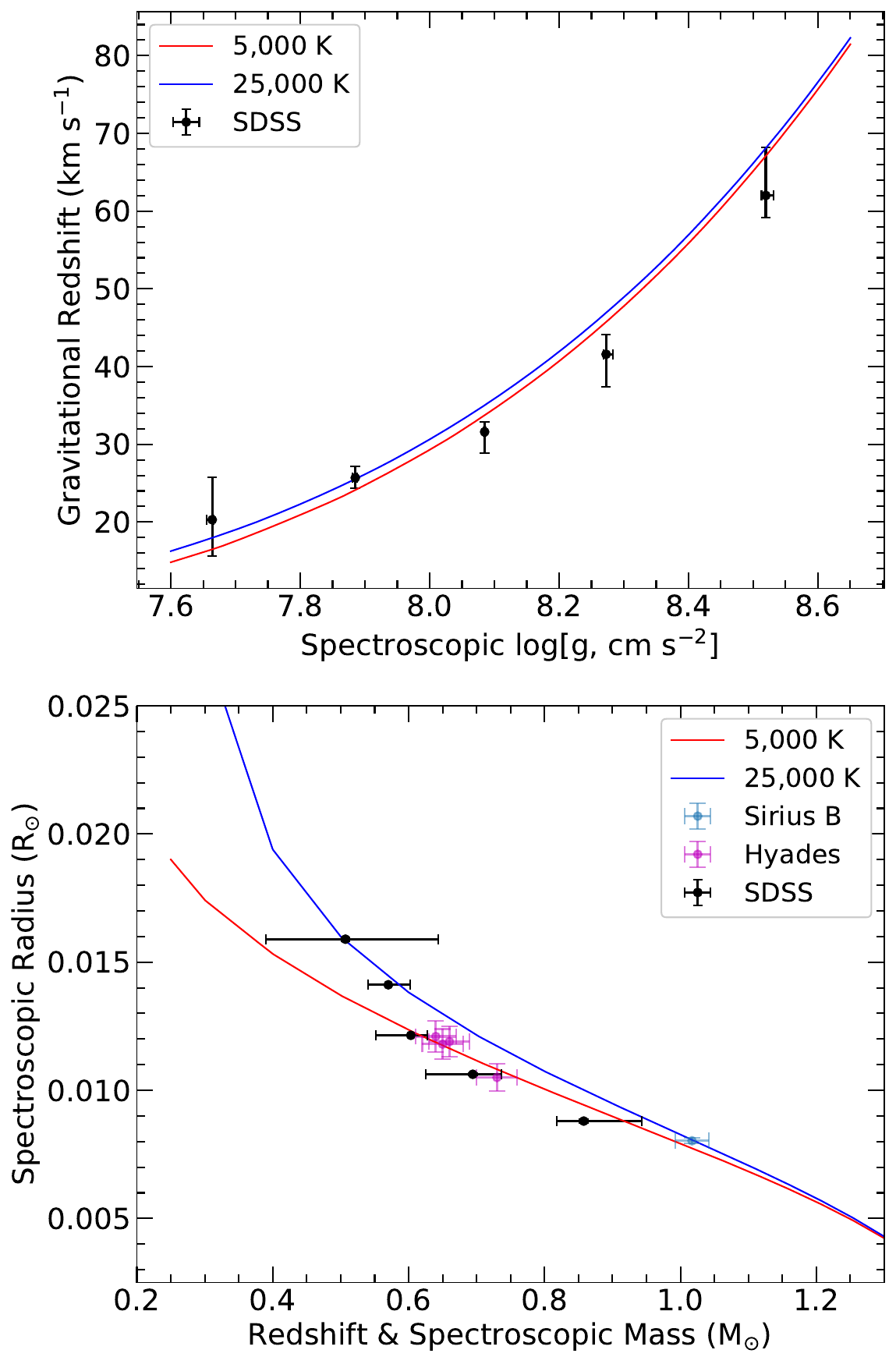}
    \caption{Gravitational redshift as a function of photometric radius (top-left) and spectroscopic surface gravity (top-right). For the photometric result we indicate our raw statistical measurements with gray circles, and our final debiased result with the black points and error bars. The bottom panels indicate derived mass--radius relations from the respective observables. We overlay theoretical models from \cite{Fontaine2001} for thick-envelope C/O cores at two temperatures and O/Ne cores at $5000\ \rm K$, along with past observations of Sirius B and the Hyades \citep{Joyce2018,Pasquini2019}. Error bars indicate 1-$\sigma$ confidence intervals.}
    \label{fig:massvrv_fontaine}\label{fig:debiased}
\end{figure*}

\subsection{Gravitational Redshift vs Radius}\label{photresult}

We bin our sample by photometric radii and compute the medians of both apparent radial velocity and photometric radii for each bin (Figure \ref{fig:massvrv_fontaine}, gray circles). Given that each bin contains at least 58 white dwarfs (and up to 1495), we assume that random stellar motion is cancelled out by this averaging, isolating the gravitational redshift \citep{Falcon2010}. We use median statistics to reduce the effect of outliers. Our photometric radius estimates are susceptible to two main biasing effects: sampling effects from the noisy white dwarf radius distribution, and the presence of unresolved binaries on the low-mass end. We estimate these bias effects with Monte Carlo simulations (see Section \ref{sec:systematic}), and present the debiased gravitational redshift as a function of radius in the upper-left panel of Figure \ref{fig:debiased} (black points with error bars).

There are two sources of uncertainty for the averaged gravitational redshift in each bin -- the intrinsic velocity dispersion from galactic kinematics (which can be as high as 10 -- 30 $\rm km\ s^{-1}$ depending on the stellar age; \citealt{Nordstrom2004}), and the uncertainty of our radial velocity measurements. Both these uncertainties are captured by the dispersion of observed radial velocities in each bin, and we therefore use the bootstrap method to derive vertical axis error bars in Figure \ref{fig:massvrv_fontaine}. We use the $16^{th}$ and $84^{th}$ percentiles of 1000 bootstrapped samples of each bin's median as the lower and upper error bars respectively, representing the 1-$\sigma$ credible interval of our statistical measurements. The bins each have (58, 188, 595, 1986, 409, 72) stars respectively. 

We transform our observables into an empirical mass--radius relation by using Equation \ref{gredshift} to convert the binned photometric radii and mean gravitational redshifts into masses. We propagate the binned redshift and radius uncertainties to derive horizontal error bars on the binned masses. This derived mass--radius relation (Figure \ref{fig:massvrv_fontaine}, bottom-left) is independent of any assumed white dwarf equation of state, and illustrates the main result of this work -- a direct statistical measurement of the white dwarf mass--radius relation across a wide range of masses. Our measurements are consistent with the theoretical mass--radius relation of \cite{Fontaine2001}. 

\subsection{Gravitational Redshift vs Surface Gravity}\label{specresult}

We repeat the statistical analysis described in Section \ref{photresult}, binning apparent radial velocities by spectroscopic $\log{g}$ and computing the median and bootstrapped error of the median. We select non-uniform bins to maintain enough stars per bin, ending up with (99, 1349, 666, 298, 165) stars in each bin. Figure \ref{fig:massvrv_fontaine} (top-right) illustrates the median gravitational redshift as a function of surface gravity. These measurements are likewise consistent with the theoretical models of \cite{Fontaine2001}.

It is difficult to derive a purely empirical mass--radius relation from our spectroscopic observables without introducing complications like correlated errors. We therefore present a semi-empirical mass--radius visualization instead. We convert spectroscopic $T_{\rm eff}$ and $\log{g}$ measurements to stellar radii by interpolating theoretical white dwarf models \citep{Fontaine2001}, and use these radii and the binned gravitational redshifts to derive masses by rearranging Equation \ref{gredshift}. We present this model-dependent visualization in Figure \ref{fig:massvrv_fontaine} (bottom-right). Uncertainties on these derived quantities are computed using Monte Carlo sampling to propagate errors from the observables.

\section{Discussion} \label{sec:discussion}
 
In this work we have empirically measured the white dwarf mass--radius relation by measuring gravitational redshift as a function of two different observables, photometric radius and spectroscopic surface gravity, using a total sample of over three thousand stars (Figure \ref{fig:massvrv_fontaine}). Within the limits of our statistical uncertainties, our results are in excellent agreement with the theoretical mass--radius relation of \cite{Fontaine2001}.  

One possible question about our analysis is whether our sample size is large enough to statistically uncover the dependence of the average gravitational redshift on radius and surface gravity. \cite{Falcon2010} used 449 DA stars to determine a single mean mass value -- our sample is almost eight times larger, albeit with noisier individual radial velocity measurements. We find that we can bin the high-precision apparent radial velocities of 449 DA stars from \cite{Falcon2010} according to their photometric radii and uncover a version of the redshift-radius relation presented in our work (similar to our Figure \ref{fig:massvrv_fontaine}, left). Due to the smaller number of stars, the redshift-radius relation we recover from their dataset is significantly noisier in every radius bin, and it does not extend over as broad a mass/radius range as ours. It is flattened by the radius measurement bias (see Section \ref{debias}) in a manner similar to our measurement. This supports our assumption that due to our much larger sample, we are not limited by uncertainties in the individual radial velocities measurements. It is more important to have accurate radius measurements to ensure that stars are placed in the correct radius bins, and to have enough stars in each bin to cancel our random stellar motions. 

As discussed in Section \ref{debias}, the main systematic uncertainty in our statistical redshift--radius relation is caused by radius uncertainties bumping stars from bins at the centre of the sharply-peaked radius distribution to other bins. We have quantified and corrected for this bias, presenting both the raw statistical measurement and the debiased relation in Figure \ref{fig:massvrv_fontaine}. The dominant uncertainty in our radius measurements is the trigonometric parallax from \textit{Gaia}, resulting in a median radius uncertainty $\sim 0.002 \ R_{\odot}$. Repeating our analysis with improved parallaxes from the upcoming \textit{Gaia} Data Release 3 should reduce the bias in our statistical redshift--radius relation. 

Looking ahead, a measurement of the core composition of white dwarfs as a function of mass would be another key probe of stellar evolution models \citep{Camisassa2019}. The top left panel of Figure \ref{fig:massvrv_fontaine} illustrates that the predicted differences between the C/O (blue and red lines) and O/Ne (orange line) compositions are small and present only for the highest masses and lowest radii. The statistical measurements in this work are not yet sensitive enough to distinguish them. However, with larger samples of white dwarfs, it may soon be possible to study core composition as a function of mass using the gravitational redshift method. Of particular interest are stars in wide astrometric binaries \citep{Tremblay2017,Joyce2018,Romero2019} where the radial velocity measurement of a co-moving companion can provide a strong constraint on the gravitational redshift of a white dwarf. \textit{Gaia} astrometry is particularly useful to search for such co-moving candidates. 
 
Another interesting line of inquiry is the dependence of the mass--radius relation on the thickness of the hydrogen layer of the atmosphere. At a fixed temperature, a thin atmosphere DA white dwarf will present with a higher spectroscopic $\log{g}$ than a thick atmosphere white dwarf of the same mass \citep{Tremblay2017}. The theoretical models used for comparison in this work assumed a `thick' hydrogen envelope with $M_{\rm H} / M_* = 10^{-4}$. However, it is known that hydrogen content can vary depending on on the mass and temperature of the white dwarf, causing a variation of up to 1-15\% in the mass--radius relation \citep{Romero2019}. In the top left panel of Figure \ref{fig:massvrv_fontaine} the variation that would be introduced by considering thick vs thin hydrogen envelopes is comparable in magnitude to the difference between the overlaid red and blue lines. With our current error bars, we cannot probe the difference introduced into the mass--radius relation by thick vs thin hydrogen layers. However, a future study with a larger sample size could be enlightening, particularly to investigate the dependence of the white dwarf hydrogen envelope on stellar mass. 

\acknowledgments

We thank the anonymous referee for their constructive feedback that significantly improved our methodology and manuscript. VC and HCH were supported by Space@Hopkins. VC was supported by the JHU PURA and DURA. VC, HCH, and NLZ were supported in part by NASA-ADAP 80NSSC19K0581. 

Based on observations from the Sloan Digital Sky Survey. Funding for the Sloan Digital Sky Survey IV \citep{SDSS2017} has been provided by the Alfred P. Sloan Foundation, the U.S. Department of Energy Office of Science, and the Participating Institutions. This work has made use of data from the European Space Agency (ESA) mission {\it Gaia} \citep{Gaia2018}, processed by the {\it Gaia} Data Processing and Analysis Consortium (DPAC). This research has made use of synthetic spectra provided by the Spanish Virtual Observatory supported from the Spanish MINECO/FEDER through grant AyA2017-84089. This research has made use of NASA’s Astrophysics Data System.

\facilities{SDSS, \textit{Gaia}}

\software{{\sc wdtools}\footnote{\url{https://github.com/vedantchandra/wdtools}} \citep{wdtools},
{\sc WD-models}\footnote{\url{https://github.com/SihaoCheng/WD_models}}, 
{\sc astropy} \citep{Astropy2013}, 
{\sc lmfit} \citep{Newville2014}, 
{\sc emcee} \citep{FM2019},
{\sc mwdust} \citep{Bovy2016},
{\sc NumPy} \citep{2011CSE....13b..22V},
{\sc SciPy} \citep{2020NatMe..17..261V},
{\sc Matplotlib} \citep{2007CSE.....9...90H}
}

\bibliography{biblio}
\bibliographystyle{aasjournal}

\end{document}